\documentclass[submission, Proceedings]{SciPost}

\pdfoutput=1

\hypersetup{
 colorlinks,
 linkcolor={red!50!black},
 citecolor={blue!50!black},
 urlcolor={blue!80!black}
}

\usepackage[bitstream-charter]{mathdesign}
\usepackage{amsmath,amsthm} 

\usepackage{enumitem} 

\newcommand{\pd}{\partial}
\renewcommand{\d}{{\operatorname{d}}}

\DeclareSymbolFont{usualmathcal}{OMS}{cmsy}{m}{n}
\DeclareSymbolFontAlphabet{\mathcal}{usualmathcal}

\begin{document}

\begin{center}{\Large \textbf{
Quantum cylindrical integrability in magnetic fields\\
}}\end{center}

\begin{center}
O. Kub\r{u}$^\star$ and L. \v{S}nobl
\end{center}

\begin{center}
Czech Technical University in Prague, Faculty of Nuclear Sciences\\ and Physical
Engineering, Prague, Czech Republic\\
* ondrej.kubu@fjfi.cvut.cz
\end{center}

\begin{center}
\today
\end{center}


\definecolor{palegray}{gray}{0.95}
\begin{center}
\colorbox{palegray}{
 \begin{tabular}{rr}
 \begin{minipage}{0.1\textwidth}
 \includegraphics[width=20mm]{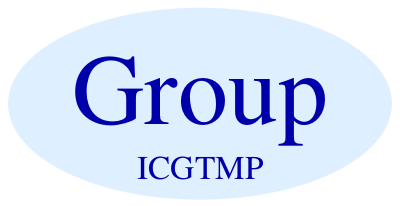}
 \end{minipage}
 &
 \begin{minipage}{0.85\textwidth}
 \begin{center}
 {\it 34th International Colloquium on Group Theoretical Methods in Physics}\\
 {\it Strasbourg, 18-22 July 2022} \\
 \doi{10.21468/SciPostPhysProc.?}\\
 \end{center}
 \end{minipage}
\end{tabular}
}
\end{center}

\section*{Abstract}
{\bf
We present the classification of quadratically integrable systems of the cylindrical type with magnetic fields in quantum mechanics. Following the direct method used in classical mechanics by [F Fournier et al 2020 J. Phys. A: Math. Theor. 53 085203] to facilitate the comparison, the cases which may \textit{a priori} differ yield 2 systems without any correction and 2 with it. In all of them the magnetic field $B$ coincides with the classical one, only the scalar potential $W$ may contain a $\hbar^2$-dependent correction. Two of the systems have both cylindrical integrals quadratic in momenta and are therefore not separable. These results form a basis for a prospective study of superintegrability.
}

\section{Introduction}
\label{sec:intro}

This article is a contribution to the study of integrable and superintegrable Hamiltonian systems with magnetic fields on the 3D Euclidean space $\mathbb{E}_3$ in quantum mechanics. More specifically, we assume a Hamiltonian of the form (using units where $e=-1,\ m=1$)
	\begin{equation}\label{HamMagn}
	H=\frac{1}{2}\left(\vec{p}^2+A_j(\vec{x})p_j+p_j A_j(\vec{x})+A_j(\vec{x})^2\right)+W(\vec{x}),
\end{equation}
with implicit summation over repeated indices $j=1,2,3$ (in the whole paper), $\vec{p}=-i\hbar\vec{\nabla}$ is the momentum operator and $\vec{A}=(A_1(\vec{x}),A_2(\vec{x}),A_3(\vec{x}))$ and $W(\vec{x})$ are the vector and scalar potentials of the electromagnetic field.

Integrability then entails the existence of two algebraically independent integrals of motion $X_1,X_2$ (further specified below) mutually in involution, i.e.
\begin{equation}\label{IC}
	[H,X_1]=[H,X_2]=[X_1,X_2]=0.
\end{equation}
They are usually considered to be polynomials in the momenta $p_j$, for computational feasibility usually of a low order (typically 2).

Integrable (and especially superintegrable) systems are rare and distinguished by the possibility to obtain the solution to their equations of motion in a closed form. They are subsequently invaluable for gaining physical intuition and serve as a starting point for modelling more complicated systems. Finding and classifying these systems is therefore of utmost importance.

The case without the vector potential $\vec{A}$ has been widely studied. The quadratic integrable systems were classified in 1960s and the 1:1 correspondence with orthogonal separation of variables of the Schr\"odinger (or, in classical context, the Hamilton-Jacobi) equation was found \cite{WinternitzSmorodinski,Makarov1967,FrisMandrosov}. This leads to the 11 classes of scalar potentials $V$ studied by Eisenhart \cite{EisenhartAnnMath}. Higher order superintegrability followed, see e.g. \cite{Escobar_Ruiz_2017} and references therein.

Despite its physical relevance, integrability with magnetic fields was mostly ignored due to its computational difficulty. The first systematic result remedying this omission was the article by Shapovalov on separable systems \cite{Shapovalov1972}, followed by the articles in $\mathbb{E}_2$ \cite{Berube,McSween2000}. Subsequent articles in $\mathbb{E}_3$ assumed first order integrals \cite{Marchesiello2015} or separation of variables \cite{Zhalij_2015,Marchesiello2018,BertrandSnobl,Fournier2019}. Marchesiello et al. \cite{Marchesiello2015} found a quadratic superintegrable system with an integral not connected to separation of variables, which was recently followed up by \cite{Marchesiello2022,Kubu2022}.

Here we present in an abridged form the classification of quadratically integrable systems of the cylindrical type (see \eqref{cyl integrals}) in quantum mechanics obtained in O. Kub\r{u}'s Master thesis \cite{Kubu2020}, which closely followed Fournier et al.'s \cite{Fournier2019} classical analysis to highlight the differences arising in quantum mechanics.

In Section \ref{sec cyl} we introduce the differential form formalism for magnetic fields in cylindrical coordinates,  derive the determining equations for cylindrical--type integrals and reduce them to a simpler form. The calculations separate into several cases depending on the rank of the matrix in equation \eqref{matrixform2}. In the case that may \textit{a priori} differ from the classical one from \cite{Fournier2019} only ranks 2 and 1 are relevant. We present the corresponding results in Sections \ref{sec:rank 2} and \ref{sec:rank 1}, respectively. We draw our conclusions in Section \ref{sec: concl}.

\section{Cylindrical--type system}\label{sec cyl}
 Before we specify the corresponding integrals $X_1, X_2$, we have to introduce the formalism used for magnetic field in curvilinear coordinates in classical mechanics, cf. \cite{Marchesiello2018Sph,Fournier2019}. 
 
 Defining the cylindrical coordinates
\begin{equation}
	 x=r \cos(\phi), \quad y=r\sin(\phi), \quad z=Z,
\end{equation}
we represent the vector potential $A$ as a 1-form
\begin{equation}
	A = A_x \mathrm{d}x + A_y \mathrm{d}y + A_z \mathrm{d}z = A_r \mathrm{d}r + A_\phi \mathrm{d}\phi + A_Z \mathrm{d}Z.
\end{equation}
Hence, we obtain the following transformations 
\begin{align}\label{transf p}
	A_x = \cos(\phi)A_r-\frac{\sin(\phi)}{r}A_\phi, \quad A_y = \sin(\phi)A_r+\frac{\cos(\phi)}{r}A_\phi, \quad A_z = A_Z.
\end{align}
As a part of the canonical 1-form $\lambda=p_j\d x^J$, the momenta $p_j$ transform in the same way and we can define the covariant momenta by $p^A_j=p_j+A_j$ in both Cartesian and cylindrical coordinates.

Components of the magnetic field 2-form $B=\mathrm{d}A$ are
\begin{align}
	\begin{aligned}
		B &= B^x (\vec{x})\,\mathrm{d}y \wedge \mathrm{d}z + B^y (\vec{x})\, \mathrm{d}z \wedge \mathrm{d}x + B^z (\vec{x})\, \mathrm{d}x \wedge \mathrm{d}y \\
		&= B^r (r, \phi, Z)\, \mathrm{d}\phi \wedge \mathrm{d}Z + B^\phi (r, \phi, Z)\, \mathrm{d}Z \wedge \mathrm{d}r + B^Z(r, \phi, Z)\, \mathrm{d}r \wedge \mathrm{d}\phi,
	\end{aligned}
\end{align}
which leads to the following transformation
\begin{align} \label{transformB}
	B^x (\vec{x}) &= \frac{\cos(\phi)}{r}B^r (r, \phi, Z) - \sin(\phi)B^\phi (r, \phi, Z), \nonumber\\
	B^y (\vec{x}) &= \frac{\sin(\phi)}{r}B^r (r, \phi, Z) + \cos(\phi)B^\phi (r, \phi, Z), \\
	B^z (\vec{x}) &= \frac{1}{r}B^Z (r, \phi, Z). \nonumber
\end{align}

We can use the same formalism and notation in quantum mechanics as well, we just have to quantize the equations and the integrals (properly symmetrized) in Cartesian coordinates and subsequently transform our equations into cylindrical ones. For example, the transformed momenta read
\begin{equation}
	 p_x=-i\hbar \left(\cos(\phi)\pd_{r}-\frac{\sin(\phi)}{r}\pd_{\phi}\right), \  p_y=-i\hbar \left(\sin(\phi)\pd_{r}+\frac{\cos(\phi)}{r}\pd_{\phi}\right), \  p_z=-i\hbar\pd_{Z},
\end{equation}
i.e. the transformation is the same as \eqref{transf p} upon defining $p_{r,\phi,Z}=-i\hbar \pd_{r,\phi,Z}$.

We can now introduce integrals of motion of the cylindrical type, i.e. integrals that imply separation of Schr\"odinger (or, classically, Hamilton-Jacobi) equation in the cylindrical coordinates in the limit of vanishing magnetic field $\vec B$. Expressed in the cylindrical coordinates they read
\begin{equation}
	\begin{split} \label{cyl integrals}
		& X_1=(p_\phi^A)^2+\frac{1}{2}\sum_{\alpha=r,\phi,Z}\left(s_1^\alpha(r, \phi, Z)p_\alpha^A+p_\alpha^A s_1^\alpha(r, \phi, Z)\right)+m_1(r,\phi,Z), \\
		& X_2=(p_Z^A)^2+\frac{1}{2}\sum_{\alpha=r,\phi,Z}\left(s_2^\alpha(r, \phi, Z)p_\alpha^A+p_\alpha^A s_2^\alpha(r, \phi, Z)\right)+m_2(r,\phi,Z).
	\end{split}
\end{equation}
The functions $s^{\{r,\phi,Z\}}_{1,2},m_{1,2}$ are to be determined from the integrability conditions \eqref{IC} together with the electromagnetic field $B,W$.

Our form of integrals allows us to separate the integrability conditions \eqref{IC} into coefficients of momenta, e.g. $p_rp_Z$, which must all vanish, yielding the so-called determining equations. The second order ones can be solved in terms of 5 auxiliary functions of one variable each, namely
\begin{align}
	\begin{split}\label{scond}
		s_1^r &=\frac{\mathrm{d}}{\mathrm{d}\phi}\psi(\phi), \quad s_1^\phi=-\frac{\psi(\phi)}{r}-r^2\mu(Z)+\rho(r), \quad s_1^Z=\tau(\phi), \\
		s_2^r&=0, \quad s_2^\phi=\mu(Z), \quad s_2^Z=-\frac{\tau(\phi)}{r^2}+\sigma(r),
	\end{split}\\
	\begin{split} \label{Bcond}
		B^r&=-\frac{r^2}{2}\frac{\mathrm{d}}{\mathrm{d}Z}\mu(Z)+\frac{1}{2r^2}\frac{\mathrm{d}}{\mathrm{d}\phi}\tau(\phi), \quad B^\phi=\frac{\tau(\phi)}{r^3}+\frac{1}{2}\frac{\mathrm{d}}{\mathrm{d}r}\sigma(r), \\
		B^Z&=\frac{-\psi(\phi)}{2r^2}+r\mu(Z)-\frac{1}{2}\frac{\mathrm{d}}{\mathrm{d}r}\rho(r)-\frac{1}{2r^2}\frac{\mathrm{d}^2}{\mathrm{d}\phi^2}\psi(\phi),
	\end{split}
\end{align}
cf. \cite{Fournier2019}. We further use primes for derivatives of these functions with respect to their variable.

We substitute this result into the remaining determining equations and the corresponding Clairaut compatibility conditions
$\pd_{ba}m_j=\pd_{ab}m_j$ and after some calculations we obtain the following reduced equations. (Indexes of $W$ mean partial derivatives.)
\begin{align}
	\psi'(\phi) \left(r^3 \sigma'(r) + 2 \tau(\phi) \right) - \tau'(\phi) \left(r \rho(r) - \psi(\phi) \right) ={}& 0, \label{reducedAa}\\
	\mu(Z) \psi'(\phi) + r^3 \sigma(r)\mu'(Z) ={}& 0, \label{reducedAb}
\end{align}
\begin{align} \label{reducedB}
	W_{r \phi} ={}&-\frac{2}{r}W_\phi + \frac{1}{4r^5}\left[\psi'(\phi) \left( r^3 (\rho''(r)- \mu(Z)) - r^2 \rho'(r) + r \rho(r) -3\psi''(\phi) - 4\psi(\phi) \right) \right. \nonumber\\
	&+ \left. \tau'(\phi) \left(r^3 \sigma'(r) + 2\tau(\phi) \right) - 2r^4 \tau(\phi) \mu'(Z) - \psi'''(\phi) \left(\psi(\phi) - r \rho(r) \right) \right], \nonumber\\
	W_{\phi Z} ={}&-\frac{1}{4r^2} \left[r^2 \mu''(Z) \left(\tau(\phi) - r^2 \sigma(r) \right) + \tau''(\phi) \mu(Z) \right], \\
	W_{r Z} ={}&\frac{1}{4r^3} \left[r \mu'(Z) \left(r^2 \rho'(r) + \psi(\phi) -2r^3 \mu(Z)\right) + 2 \mu(Z) \tau'(\phi) \right], \nonumber
\end{align}

\begin{align} \label{matrixform2}
	\begin{pmatrix}
		0 & r^2 \mu(Z) & r^2 \sigma(r) - \tau(\phi)\\
		\psi'(\phi) & \rho(r) -r^2 \mu(Z) - \frac{\psi(\phi)}{r} & \tau(\phi)\\
		0 & 4 r^7 \mu(Z) & -4 r^5 \tau(\phi)
	\end{pmatrix}
	\cdot \begin{pmatrix}
		W_r\\
		W_\phi\\
		W_Z
	\end{pmatrix}=
	\begin{pmatrix}
		0\\
		-\frac{\hbar^2(\psi'''(\phi)+\psi'(\phi))}{4r^3}\\
		0
	\end{pmatrix}.
\end{align}
We denote the matrix in \eqref{matrixform2} by $M$.

The only change with respect to the classical case is the non-zero RHS in equation \eqref{matrixform2}, corresponding to equation (39) in \cite{Fournier2019}.

We proceed depending on the rank of the matrix $M$: rank 0 implies vanishing magnetic field and rank 3 is inconsistent with the other reduced equations \eqref{reducedAa}--\eqref{reducedB}, we therefore consider ranks 1 and 2 only.

For both these ranks we have to consider the following 2 cases,
\begin{enumerate}[label=\alph*)]
	\item $\psi'(\phi)=0$ \label{case a},
	\item $\psi'(\phi) \neq 0$ and $\mu(Z)=0$\label{case b}.
\end{enumerate}

Case \ref{case a} implies vanishing quantum correction in \eqref{matrixform2}, i.e. the systems are characterized by the same functions $B, W, s$ and $m$ in the classical and quantum mechanics. The reader can find those in \cite{Fournier2019}. Hereafter we present only the results for case \ref{case b} for brevity, details can be found in \cite{Kubu2020}.

\section{$\mathrm{rank}\, (M)=2$}\label{sec:rank 2}
Here the matrix equation \eqref{matrixform2} implies that the coordinate $Z$ is cyclical in a suitable gauge and the integral $X_2$ reduces to a first order one $p_Z=p_z$ in that gauge.

We obtain 2 systems. The dynamics of the first one splits to a free motion in the $z$ direction and a 2D system with perpendicular magnetic field, so some details were extracted from the consideration of 2D systems in \cite{Berube,McSween2000}. The other system cannot be separated in this way due to a more complicated magnetic field (but $p_z$ remains an integral).

In what follows we use $r=\sqrt{x^2+y^2}$ for brevity in Cartesian coordinates as well and expressions like $\rho_{1},\psi_{2}$ and $W_0$ are (usually nonvanishing) constants unless a variable is explicitly indicated.
\begin{enumerate}
			\item\label{i} The first system is classical, i.e. the quantum correction vanishes. The electromagnetic field in Cartesian coordinates reads
			\begin{equation}\label{11a}
				\begin{split}
					B^x={}&0,\quad B^y=0,\quad B^z=-6\rho_2 r^2+\rho_1,\\
					W={}&-2\rho_2 (\psi_1 x+\psi_2 y)-\rho_2^2 r^6+
					\frac{\rho_2 \rho_1}{2}r^4-\rho_2 W_0r^2.\raisetag{2\baselineskip}
				\end{split}
			\end{equation}
			
			The cylindrical integral of motion $X_1$ in Cartesian coordinates is
			\begin{flalign}
				\begin{aligned}
					X_1={}&(L_z^A)^2+(3\rho_2 r^4-\rho_1 r^2+W_0)L_z^A-\psi_2 p_x^A+\psi_1 p_y^A\\
					&+(2\rho_2 r^2-\rho_1 )( \psi_1x+\psi_2y)\\
					&+\tfrac{1}{4}(3 \rho_2 r^4 - \rho_1 r^2 + 2 W_0) (3 \rho_2 r^2- \rho_1) r^2,
				\end{aligned}
			\end{flalign}	
		
		\item\label{1.2}
		This time the reduced equations  \eqref{reducedAa}--\eqref{matrixform2} were not completely solved.
		The electromagnetic field listed below features a function $\beta(\phi)=\psi(\phi)-\rho_0$ which must satisfy the following ODE
		\begin{align} \label{betasimple}
			\beta'(\phi) \left( 7 \beta(\phi)\beta''(\phi)+4\beta'(\phi)^2 + 12 \beta(\phi)^2 + f_1 \right) + \beta(\phi)^2\beta'''(\phi)=0,
		\end{align}
	where $f_1$ is a parameter, or equivalently its reduced form with integrating constants $\beta_1,\beta_2$
		\begin{equation} \label{betafirstorder}
			4 \beta(\phi)^4 \beta'(\phi)^2 + 4 \beta(\phi)^6 - 4 \beta_1 \beta(\phi)^2+ f_1 \beta(\phi)^4 = \beta_2.
		\end{equation} 
	We were not able to solve this autonomous first order differential equation in general in an explicit form. As was noted in \cite{Berube}, a hodograph transformation leads to $\phi(\beta)$ obtainable by a quadrature in terms of elliptic integrals. However, the subsequent inversion to find $\beta(\phi)$ is either impossible or not illuminating, therefore the authors of \cite{Berube} proceed only with some special solutions. e.g. by choosing $\beta_2=0$, $f_1<0$, and $-\frac{f_1^2}{64}<\beta_1<0$, we obtain a well-defined solution
	\begin{equation}\label{beta beta2=0}
		\beta(\phi)=\sqrt{\frac{\sqrt{64 \beta_1+f_1^2}\sin\left( 2 (\phi-\phi_0) \right)-f_1}{8}}.
	\end{equation}

	Using the ODEs above to eliminate derivatives of $\beta(\phi)$, the electromagnetic field for any solution $\beta(\phi)$ of \eqref{betafirstorder} is given by
		\begin{align}\label{12B}
	B^r ={}&-\tau_1 \frac{\sqrt{4 \beta_1 \beta(\phi)^2+\beta_2-4\beta(\phi)^6- f_1 \beta(\phi)^4}}{2 r^2 \beta(\phi)^5}, \nonumber\\
	B^\phi ={}&\frac{\tau_1}{r^3 \beta(\phi)^2}, \quad B^Z = \frac{2 \beta_1 \beta(\phi)^2+\beta_2}{4 r^2 \beta(\phi)^5},\\
	W ={}& \frac{W_0}{r^2 \beta(\phi)^2}-\frac{(4 \tau_1^2 + \beta_2)}{32 r^4 \beta(\phi)^4}
	+\hbar^2\frac{f_1 \beta(\phi)^4 -12 \beta_1 \beta(\phi)^2 - 5 \beta_2}{32 r^2 \beta(\phi)^6}\nonumber.
\end{align}
		The sign of the square root depends on the branch chosen while substituting for $\beta'(\phi)$ from \eqref{betafirstorder}.
		
		We note that the magnetic field is the same classically and quantum mechanically in both cases, only the scalar potential $W$ obtains an $\hbar^2$-proportional correction depending on $\beta$. This system admits $\tau_1=0$, which leads to its separation into free 1D motion plus 2D motion in a perpendicular magnetic field.
		
		The lower order terms of the cylindrical integral $X_1 $ from \eqref{cyl integrals} are determined by
		\begin{equation}
		\begin{split}
		s_1^r ={}&\frac{\sqrt{4 \beta_1 \beta(\phi)^2+\beta_2-4 \beta(\phi)^6-\beta(\phi)^4 f_1}}{2 \beta(\phi)^2},\\
		s_1^\phi ={}&-\frac{\beta(\phi)}{r}, \qquad s_1^Z = \tau_0+\frac{\tau_1}{\beta(\phi)^2}, \\
		m_1 ={}&\frac{2 W_0}{\beta(\phi)^2}-\frac{4 \beta(\phi)^2 \tau_0 \tau_1+2 \beta_1 \beta(\phi)^2+4 \tau_1^2+\beta_2}{8 \beta(\phi)^4 r^2}\\
		&+\hbar^2\frac{f_1 \beta(\phi)^4 -12 \beta_1 \beta(\phi)^2 - 5 \beta_2}{16 r^2 \beta(\phi)^6},
	\end{split}
		\end{equation}
	
		the integral $X_2$ reduces to the first order one $\tilde{X}_2=p_Z$ in a suitable gauge.
	\end{enumerate}

\section{$\mathrm{rank}\, (M)=1$}\label{sec:rank 1}

Here the general form of the fields is as follows
			\begin{equation}
	\begin{split}
		B^x={}&0,\quad B^y=0,\quad B^z=B^z(x,y),\\
		W={}&W_{12}(x,y)+W_3(z).
	\end{split}
\end{equation}

This implies that the system again separates into the 1D motion in the $z$ direction, influenced by the scalar potential $W_3(z)$ and no magnetic field ($B^x=B^y=0$), and the motion in the $xy$ direction determined by a perpendicular magnetic field $B^z(x,y)$ and a scalar potential $W_{12}(x,y)$ containing \textit{a priori} a quantum correction.

The presence of the scalar potential $W_3(z)$, which is not constrained further, implies that the cylindrical integral,
\begin{equation}
	X_2=(p_z^A)^2+2W_{3}(z),
\end{equation}
does not reduce to a first order one and the motion in the $z$ direction is no longer free.

The remaining 2D motion, studied earlier in \cite{Berube,McSween2000}, is integrable due to the integral $X_1$ from \eqref{cyl integrals}. The relevant magnetic field $B^z(x,y)$, the 2D scalar potential $W_{12}(x,y)$ and the functions $s^{r,\phi,Z}_1,m_1$ coincide with the results of Section \ref{sec:rank 2}, namely case \ref{i}, see \eqref{11a}, and case \ref{1.2} with $\tau_1=0$, see \eqref{12B}. In both cases neither of the cylindrical integrals reduces to a first order one (with potential exceptions for some special solutions of $\beta(\phi)$), therefore the results of \cite{Shapovalov1972,Benenti_2001} imply that these systems are in general not separable.

\section{Conclusions}\label{sec: concl}
In this article we presented the classification of quantum quadratically integrable systems of cylindrical type obtained in O. Kub\r{u}'s Master thesis \cite{Kubu2020}. We proceeded by directly solving the determining equations \eqref{IC}. Despite our focus on the quantum case, we followed the analysis from the classical case \cite{Fournier2019} to facilitate comparison and because only the zeroth order equations contain a correction, see the reduced equation \eqref{matrixform2}. 

Noting that the results of case \ref{case a} coincide with the classical ones known from \cite{Fournier2019}, we analyze further only case \ref{case b} where the quantum correction is \emph{a priori} non--trivial. We find that in all remaining subcases the magnetic fields coincide with their classical counterparts and only the scalar potential $W$ is modified by a $\hbar^2$-proportional correction. However, even here it may vanish due to the consistency conditions on the scalar potential $W$, leaving us with 2 systems with a correction, namely system \ref{1.2} in Section \ref{sec:rank 2} and its counterpart in Section \ref{sec:rank 1} (in the latter $\tau_1=0$ is necessary).

In all cases there is at least one free parameter, a constant or even a function. It is therefore probable that some superintegrable systems can be found by imposing further restrictions. This has been done in classical mechanics for separable systems \cite{Kubu_2021} and on the intersection with other integrable systems \cite{Bertrand2020}, but only for first order integrals in quantum mechanics \cite{Kubu2020}. Easing these restrictions is necessary as well as going beyond integrals connected to orthogonal separation of variables as was shown in \cite{Marchesiello2022}.

\section*{Acknowledgements}
The research in OK's Master thesis, on which the article is based, was supported by the Czech Science Foundation
(Grant Agency of the Czech Republic), project 17-11805S, and Grant Agency of the Czech Technical University in
Prague, grant No. SGS19/183/OHK4/3T/14. OK thanks late Pavel Winternitz, his supervisor and host during his ERASMUS+ stay at Universit\'{e} de Montr\'{e}al, where the work on his Master thesis started. OK's presentation at Group 34 conference and recent research is supported by the Grant Agency of the Czech Technical University in Prague, grant No. SGS22/178/OHK4/3T/14.



\bibliography{GROUP34.bib}

\begin{thebibliography}{10}
\providecommand{\url}[1]{\texttt{#1}}
\providecommand{\urlprefix}{URL }
\expandafter\ifx\csname urlstyle\endcsname\relax
  \providecommand{\doi}[1]{doi:\discretionary{}{}{}#1}\else
  \providecommand{\doi}{doi:\discretionary{}{}{}\begingroup
  \urlstyle{rm}\Url}\fi
\providecommand{\eprint}[2][]{\url{#2}}

\bibitem{WinternitzSmorodinski}
P.~Winternitz, J.~A. Smorodinsky, M.~Uhl\'{\i}\v{r} and I.~Fri\v{s},
\newblock \emph{Symmetry groups in classical and quantum mechanics},
\newblock Soviet J. Nuclear Phys. \textbf{4}, 444 (1967),
\newblock \doi{10.1007/s002200100454}.

\bibitem{Makarov1967}
A.~Makarov, J.~Smorodinsky, K.~Valiev and P.~Winternitz,
\newblock \emph{A systematic search for nonrelativistic systems with dynamical
  symmetries},
\newblock Nuovo Cimento A Series \textbf{10}, 1061 (1967),
\newblock \doi{10.1007/BF02755212}.

\bibitem{FrisMandrosov}
J.~Fri\v{s}, V.~Mandrosov, J.~A. Smorodinsky, M.~Uhl\'{\i}\v{r} and
  P.~Winternitz,
\newblock \emph{On higher symmetries in quantum mechanics},
\newblock Phys. Lett. \textbf{16}, 354 (1965),
\newblock \doi{10.1016/0031-9163(65)90885-1}.

\bibitem{EisenhartAnnMath}
L.~P. Eisenhart,
\newblock \emph{Separable systems of {S}tackel},
\newblock Ann. of Math. (2) \textbf{35}(2), 284 (1934),
\newblock \doi{10.2307/1968433}.

\bibitem{Escobar_Ruiz_2017}
A.~M. Escobar-Ruiz, J.~C. L{\'{o}}pez~Vieyra and P.~Winternitz,
\newblock \emph{Fourth order superintegrable systems separating in polar
  coordinates. {I}. {E}xotic potentials},
\newblock J. Phys. A: Math. Theor. \textbf{50}(49), 495206 (2017),
\newblock \doi{10.1088/1751-8121/aa9203}.

\bibitem{Shapovalov1972}
V.~N. Shapovalov, V.~G. Bagrov and A.~G. Meshkov,
\newblock \emph{Separation of variables in the stationary {S}chroedinger
  equation},
\newblock Soviet Physics Journal \textbf{15}(8), 1115 (1972),
\newblock \doi{10.1007/bf00910289}.

\bibitem{Berube}
J.~B{\'{e}}rub{\'{e}} and P.~Winternitz,
\newblock \emph{Integrable and superintegrable quantum systems in a magnetic
  field},
\newblock J. Math. Phys. \textbf{45}(5), 1959 (2004),
\newblock \doi{10.1063/1.1695447}.

\bibitem{McSween2000}
E.~McSween and P.~Winternitz,
\newblock \emph{Integrable and superintegrable {H}amiltonian systems in
  magnetic fields},
\newblock J. Math. Phys. \textbf{41}(5), 2957 (2000),
\newblock \doi{10.1063/1.533283}.

\bibitem{Marchesiello2015}
A.~Marchesiello, L.~\v{S}nobl and P.~Winternitz,
\newblock \emph{Three-dimensional superintegrable systems in a static
  electromagnetic field},
\newblock J. Phys. A \textbf{48}(39), 395206, 24 (2015),
\newblock \doi{10.1088/1751-8113/48/39/395206}.

\bibitem{Zhalij_2015}
A.~Zhalij,
\newblock \emph{Quantum integrable systems in three-dimensional magnetic
  fields: the {C}artesian case},
\newblock J. Phys.: Conf. Ser. \textbf{621}, 012019 (2015),
\newblock \doi{10.1088/1742-6596/621/1/012019}.

\bibitem{Marchesiello2018}
A.~Marchesiello and L.~\v{S}nobl,
\newblock \emph{An infinite family of maximally superintegrable systems in a
  magnetic field with higher order integrals},
\newblock SIGMA Symmetry Integrability Geom. Methods Appl. \textbf{14}, 092, 11
  (2018),
\newblock \doi{10.3842/SIGMA.2018.092}.

\bibitem{BertrandSnobl}
S.~Bertrand and L.~\v{S}nobl,
\newblock \emph{On rotationally invariant integrable and superintegrable
  classical systems in magnetic fields with non-subgroup type integrals},
\newblock J. Phys. A \textbf{52}(19), 195201, 25 (2019),
\newblock \doi{10.1088/1751-8121/ab14c2}.

\bibitem{Fournier2019}
F.~Fournier, L.~\v{S}nobl and P.~Winternitz,
\newblock \emph{Cylindrical type integrable classical systems in a magnetic
  field},
\newblock J. Phys. A: Math. Theor. \textbf{53}(8), 085203 (2020),
\newblock \doi{10.1088/1751-8121/ab64a6}.

\bibitem{Marchesiello2022}
A.~Marchesiello and L.~{\v{S}}nobl,
\newblock \emph{Pairs of commuting quadratic elements in the universal
  enveloping algebra of {E}uclidean algebra and integrals of motion},
\newblock J. Phys. A: Math. Theor. \textbf{55}(14), 145203 (2022),
\newblock \doi{10.1088/1751-8121/ac515e}.

\bibitem{Kubu2022}
O.~Kub\r{u}, A.~Marchesiello and L.~\v{S}nobl,
\newblock \emph{New classes of quadratically integrable systems in magnetic
  fields: the generalized cylindrical and spherical cases}  (2022),
\newblock \eprint{https://arxiv.org/abs/2206.15305}.

\bibitem{Kubu2020}
O.~Kub\r{u},
\newblock \emph{Integrable and superintegrable systems of cylindrical type in
  magnetic fields}  (2020),
\newblock Master's thesis, Czech Technical University in Prague,
\newblock \eprint{https://arxiv.org/abs/2210.02393}.

\bibitem{Marchesiello2018Sph}
A.~Marchesiello, L.~\v{S}nobl and P.~Winternitz,
\newblock \emph{Spherical type integrable classical systems in a magnetic
  field},
\newblock J. Phys. A \textbf{51}(13), 135205, 24 (2018),
\newblock \doi{10.1088/1751-8121/aaae9b}.

\bibitem{Benenti_2001}
S.~Benenti, C.~Chanu and G.~Rastelli,
\newblock \emph{Variable separation for natural {H}amiltonians with scalar and
  vector potentials on {R}iemannian manifolds},
\newblock J. Math. Phys. \textbf{42}(5), 2065 (2001),
\newblock \doi{10.1063/1.1340868}.

\bibitem{Kubu_2021}
O.~Kub{\r{u}}, A.~Marchesiello and L.~{\v{S}}nobl,
\newblock \emph{Superintegrability of separable systems with magnetic field:
  the cylindrical case},
\newblock J. Phys. A: Math. Theor. \textbf{54}(42), 425204 (2021),
\newblock \doi{10.1088/1751-8121/ac2476}.

\bibitem{Bertrand2020}
S.~Bertrand, O.~Kub{\r{u}} and L.~{\v{S}}nobl,
\newblock \emph{On superintegrability of 3{D} axially-symmetric
  non-subgroup-type systems with magnetic fields},
\newblock J. Phys. A: Math. Theor. \textbf{54}(1), 015201 (2020),
\newblock \doi{10.1088/1751-8121/abc4b8}.

\end{thebibliography}

\nolinenumbers

\end{document}